\documentclass{ws-modza}

\begin{document}

\title{Local Energy--Density Functional Approach to\\
Many--Body Nuclear Systems with S--Wave Pairing}

\author{S. A. Fayans}

\address{Russian Research Centre -- ``Kurchatov Institute'', 123182, 
Moscow, Russia \\ E-mail: fayans@mbslab.kiae.ru}

\author{D. Zawischa}

\address{Institut f\"ur Theoretische Physik, Universit\"at Hannover,
D-30060 Hannover, Germany\\
E-mail: zawischa@itp.uni-hannover.de}  

\maketitle

\abstracts{The ground-state properties of superfluid nuclear systems with 
$^1S_0$ pairing are studied within a local energy-density functional (LEDF) 
approach. A new form of the LEDF is proposed with a volume part which fits 
the Friedman-Pandharipande and Wiringa-Fiks-Fabrocini equation of state 
at low and moderate densities and allows an extrapolation to higher 
densities preserving causality. For inhomogeneous systems, a surface term is 
added, with two free parameters, which has a fractional form like a Pad\'e 
approximant containing the gradient of density squared in both the numerator 
and denominator. In addition to the Coulomb direct and exchange interaction 
energy, an effective density-dependent Coulomb-nuclear correlation term is 
included with one more free parameter. A three-parameter fit to the masses 
and radii of about 100 spherical nuclei has shown that the latter term gives 
a contribution of the same order of magnitude as the Nolen-Schiffer anomaly in 
Coulomb displacement energy. The root-mean-square deviations from experimental 
masses and radii with the proposed LEDF come out about a factor of two smaller 
than those obtained with the conventional functionals based on the Skyrme or 
finite-range Gogny force, or on the relativistic mean-field theory. 
The generalized variational principle is formulated leading to the 
self-consistent Gor'kov equations which are solved exactly, with physical 
boundary conditions both for the bound and scattering states.
The method is used to calculate the differential observables such as 
odd-even mass differences and staggering in charge radii. With a zero-range 
density-dependent cutoff pairing interaction incorporating a 
density-gradient term, the evolution of these observables is reproduced 
reasonably well, including kinks at magic neutron numbers and sizes of 
staggering. An extrapolation from the pairing properties 
of finite nuclei to pairing in infinite nuclear matter is discussed. A 
``reference'' value of the pairing gap $\Delta_F \approx 3.3$~MeV is 
found for subsaturated nuclear matter at about 0.65 of the equilibrium 
density. With the formulated LEDF approach, we study also the dilute limit 
in both the weak and strong coupling regime. Within the sum rules approach 
it is shown that the density-dependent pairing may also induce sizeable 
staggering and kinks in the evolution of the mean energies of 
multipole excitations.}

\section{Introduction}
The challenge of deriving the properties of nuclear matter and finite nuclei 
starting with a ``realistic'' bare NN interaction stimulates broad 
activity in developing the microscopic approaches to strongly coupled
many-body fermion systems. Significant progress in reproducing the empirical 
data has been made with the quantum Monte Carlo calculations for light 
nuclei\cite{W98} and with the variational chain summations method for  
nucleon matter\cite{APR98}. However, to ensure the proper binding energy of
finite nuclei and the equilibrium properties of nuclear matter, these
approaches make use of some plausible density-dependent or three-nucleon 
phenomenological terms which are additionally included in the nuclear 
Hamiltonian. Then, with the present-day Monte Carlo calculations, an accuracy 
within 1\% could be obtained for the binding energies of light $p$-shell 
nuclei up to the mass number $A=8$, but, despite the rapid growth of 
computational power, for heavier nuclei such calculations are not feasible in 
nearest future. Moreover, unfortunately, this kind of {\em ``ab initio''\/} 
approaches (like the Brueckner-Bethe-Goldstone many-body theory, see e.g. 
Ref.\cite{BBB97}) do not yet provide an effective interaction, or nuclear 
energy-density functional, which could be used in nuclear structure 
calculations with sufficient accuracy to meet modern experiments and to give 
reliable predictions for nuclei far from the measured region. An important 
issue is how to reveal the density dependence of the effective 
interaction particularly in the particle-particle (pp) channel where the 
$s$-wave pairing correlations are known to play an essential role in finite 
nuclei.

Among the existing {\it effective} models, the most successful are 
the phenomenological mean-field microscopic approaches incorporating 
density-dependent forces of the Skyrme type with zero range or of the Gogny 
type with finite range, and also the relativistic mean field (RMF) model with 
classical meson fields. With some 10 fitting parameters, these models 
can give the masses and radii of measured nuclei with the respective rms 
deviations $\approx 2$~MeV and $\approx 0.02$~fm from 
experiment\cite{Pom97,Pat99}. At the same time, they reveal a large spread
in the extrapolation behavior to the unexplored regions of the nuclear chart
and their predictions, already for nuclei not too far from stability, often 
deviate significantly from those of the macroscopic-microscopic (MM) 
models\cite{MM} or of the extended Thomas-Fermi model with Strutinsky integral 
(ETFSI)\cite{ETFSI}. These latter models are able to reproduce the 
masses and charge radii of known nuclei with rms errors down to 
$\approx 0.6$~MeV and $\approx 0.02$~fm, respectively. One may still notice 
that these errors are not mutually consistent: the relative rms deviations 
for radii are a few times larger than those for masses. Now, the description 
of the bulk nuclear properties within a microscopic model with the accuracy 
on the level of about 0.1\% would be considered as quite successful. 

\section{A new form of the LEDF}
Recently, a new {\it ansatz} for the LEDF construction has been 
attempted\cite{Fa98} to diminish the gap between the predictions of the fully 
self-consistent microscopic mean-field models and the quasi-classical or MM 
models, hunting also for a more universal nuclear density functional 
which could be used not only throughout the nuclear chart but also for
describing such objects as neutron stars, with crystal structure in their 
crust. The energy density of a nuclear system is represented as
\begin{equation}
\varepsilon =
\varepsilon_{\rm kin}+\varepsilon_{\rm v} + \varepsilon_{\rm s} +
\varepsilon_{\rm Coul} + \varepsilon_{sl} + \varepsilon_{\rm anom}~,
\label{eden}
\end{equation}
where $\varepsilon_{\rm kin}$ is the kinetic energy term which, since
we are constructing a Kohn-Sham type functional, is taken with the free
operator $t=p^2/2m$, i.e. with the effective mass $m^*=m$; all the other
terms are discussed below.

\begin{figure}[b]
\begin{center}
\epsfxsize=20pc
\epsfbox{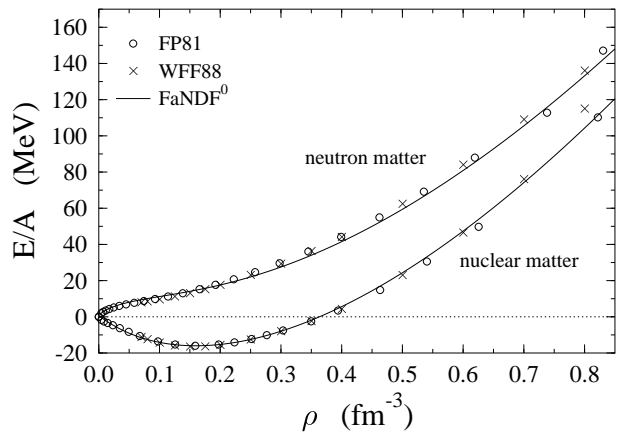}
\end{center}
\caption{The energy per nucleon versus density in nuclear and 
neutron matter. Open circles and crosses are the 
calculations by Friedman and Pandharipande$^{10}$      
and Wiringa {\it et al.}$^{11}$,                       
respectively. The solid curves are obtained from Eq.~(\ref{evol}) with the 
parameters given in the text. \label{fig:eos}}
\end{figure}

The volume term in~(\ref{eden}) is chosen to be in the form
\begin{equation}
\varepsilon_{\rm v} = {2\over 3}\epsilon_{\rm F}^0\rho_0 \left[
a^{\rm v}_+ {{1-h_{1+}^{\rm v} x_+^\sigma}\over{1+h_{2+}^{\rm v} 
x_+^\sigma}} x_+^2 +
a^{\rm v}_- {{1-h_{1-}^{\rm v} x_+}       
\over{1+h_{2-}^{\rm v} x_+       }} x_-^2
\right]\,. \label{evol}
\end{equation}
Here and in the following $x_{\pm} = (\rho_{\rm n}\pm\rho_{\rm p})/2\rho_0$ 
with $\rho_{\rm n(p)}$ the neutron (proton) density, $2\rho_0$ is the 
equilibrium density of symmetric nuclear matter with 
$\epsilon_{\rm F}^0=(9\pi/8)^{2/3}\hbar^2/2mr_0^2$, the Fermi energy
and $r_0=(3/8\pi\rho_0)^{1/3}$, the radius parameter. The fractional
expressions of the type of Eq.~(\ref{evol}) were introduced in 
Ref.\cite{STF88}
for the LEDF with application to finite systems with $s$-wave pairing 
correlations. Such expressions allow an extrapolation of the nuclear equation 
of state (EOS) to very high densities while preserving causal behavior. This 
might be of advantage since the available microscopic EOS often violate 
causality at $\rho > 1$~fm$^{-3}$. Thus, in deriving the parameters of 
Eq.~(\ref{evol}), we shall use the EOS of Refs.\cite{FP81,WFF88} only in the 
region of up to about six times the saturation density. The four parameters 
in the isoscalar volume energy density $\propto\!a_+^{\rm v}$ are found by 
fitting 
the EOS of symmetric infinite nuclear matter\cite{FP81,WFF88} for the UV14 
plus TNI model. The result shown in Fig.~\ref{fig:eos} by the lower solid 
curve is obtained with the exponent $\sigma=1/3$, the compression modulus 
$K_0=220$~MeV, the equilibrium density $2\rho_0=0.16$~fm$^{-3}$ 
($r_0=1.143$~fm) and the chemical potential $\mu=-16.0$~MeV (the energy per 
nucleon at saturation point). The corresponding dimensionless parameters are 
$a_+^{\rm v}=-9.559,\,h_{1+}^{\rm v}=0.633,\,h_{2+}^{\rm v}=0.131$. 
Keeping them fixed, a fit 
to the neutron matter EOS from the same papers\cite{FP81,WFF88} is performed 
to determine the three parameters of the isovector 
part $\propto\!a^{\rm v}_-$ in 
Eq.~(\ref{evol}). The result shown in Fig.~\ref{fig:eos} by the upper solid 
curve is obtained with 
$a_-^{\rm v}=4.428,\,h^{\rm v}_{1-}=0.250,\,h^{\rm v}_{2-}=0.130$. This 
corresponds to the asymmetry energy coefficient $\beta_0=30.0$~MeV.
The surface part $\varepsilon_{\rm s}$
in Eq.~(\ref{eden}) is meant to describe the finite-range
and nonlocal in-medium effects which, phenomenologically, may presumably be
incorporated within the LEDF framework in a localized form by introducing a
dependence on density gradients. It is taken as follows:
\begin{equation}
\varepsilon_{\rm s} = {2\over 3}\epsilon_{\rm
F}^0\rho_0 {{a_+^{\rm s} r_0^2 (\vec \nabla x_+)^2}   \over {1+h^{\rm s}_+
x_+^\sigma+h^{\rm s}_\nabla r_0^2(\vec \nabla x_+)^2}}\,, \label{esurf}
\end{equation}
with $h^{\rm s}_+=h^{\rm v}_{2+}$, and the two free parameters
$a^{\rm s}_+$ and $h^{\rm s}_\nabla$. This peculiar surface term may be 
regarded as the Pad\'e approximant for the (unknown) expansion in 
$(\vec \nabla \rho)^2/(1+h_+^{\rm s} x_+^\sigma)$ where the form factor 
$1/(1+h_+^{\rm s} x_+^\sigma)$ imitates a transformation to Migdal's 
quasiparticles (cf. Ref.\cite{KSap}). In fact, $h_+^{\rm s}$ is also a 
free parameter but here we prefer to keep it fixed by the above condition.

The Coulomb part in Eq.~(\ref{eden}) is approximated by
\begin{equation}
\varepsilon_{\rm Coul}=\frac 1 2 e^2\rho_{\rm ch}(\vec r)
\int \frac{\rho_{\rm ch}(\vec r\,')d\vec r\,'}
{\vert\vec r - \vec r\,'\vert}
-{3\over 4}({3\over \pi})^
{1/3}e^2\rho_{\rm p}^{4/3}(1-h_{\rm Coul}x_+^\sigma)\,, \label{eCoul}
\end{equation}
where the first term is the direct Coulomb contribution expressed through 
charge density $\rho_{\rm ch}$ while the second term is the exchange part 
taken in the Slater approximation and combined with the Coulomb-nuclear 
correlation term $\propto\!h_{\rm Coul}$. The latter is believed to account 
for the correlated motion of protons in nuclei beyond the direct (Hartree)
and exchange (Fock) Coulomb interaction\cite{BSh96}. The parameter
$h_{\rm Coul}$ allows to practically kill one more enemy: the Nolen-Schiffer 
anomaly.

The spin--orbit term $\varepsilon_{sl}$ in Eq.~(\ref{eden}) is related to
the two-body spin-orbit interaction $\propto\!(\kappa +
\kappa'\vec\tau_1\cdot\vec\tau_2)[ {\vec \nabla}_1 \delta({\vec
r}_1-{\vec r}_2)\times({\vec p}_1- {\vec p}_2)]\cdot(\vec\sigma_1+
\vec\sigma_2)$. It is known from the RMF theory that the isovector spin-orbit 
force is very small compared to the isoscalar one. Thus we set 
$\kappa'=0$ and derive the isoscalar strength $\kappa=0.19$ from the average 
description of the splitting of the single-particle states in $^{208}$Pb.

The last term in Eq.~(\ref{eden}), the anomalous energy density,
is represented as
\begin{equation}
\varepsilon_{\rm anom} = \sum_{i={\rm n,p}}
\nu^i({\vec r}){\cal F}^\xi(x_+({\vec r}))
\nu^i({\vec r})\,, \label{eanomal}
\end{equation}
where $\nu({\vec r})$ is the anomalous density and 
${\cal F}^\xi=C_0f^{\xi}$ is
the effective force in the particle--particle channel with the dimensionless
form factor\cite{FZ96}
\begin{equation} 
f^\xi(x_+) = f^\xi_{\rm ex} +h^\xi x_+^q + f^\xi_\nabla
r_0^2({\vec \nabla}x_+)^2\,. \label{fxi} 
\end{equation}
Here $C_0=2\epsilon^0_{\rm F}/3 \rho_0$ is the inverse density of states on 
the Fermi surface ($C_0=307.2$~MeV$\cdot$fm$^3$); $q=1$. The 
strength parameters $f^\xi_{\rm ex}=-2.8$, $h^\xi=2.8$ and $f^\xi_\nabla=2.2$ 
are extracted from a fit to the neutron separation energies and charge radii 
of lead isotopes\cite{FZ96}.
\begin{figure}[t]
\begin{center}
\epsfxsize=20pc
\epsfbox{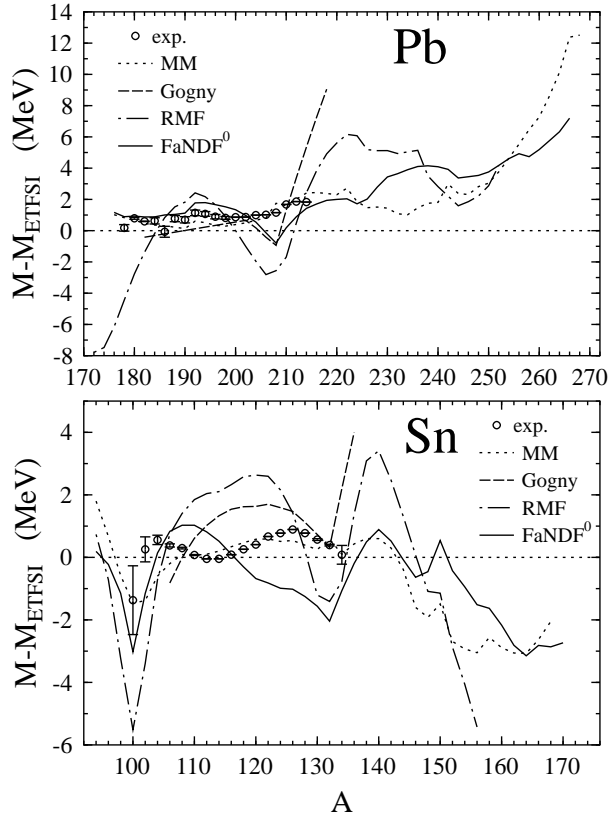}
\end{center}
\caption{Deviations of various theoretical masses from the ETFSI mass$^7$ 
for lead and tin isotopes. 
Dotted lines correspond to the MM model$^6$,         
dashed to the Gogny force$^5$                        
and dash-dotted to the RMF model$^{15}$.             
The results obtained with FaNDF$^0$ are shown by solid lines. The 
experimentally known masses$^{16}$
are presented by open circles. \label{fig:pbsn}}
\end{figure}

The three parameters $a^{\rm s}_+$, $h^{\rm s}_\nabla$ and $h_{\rm Coul}$ 
remain to be defined. This was done through a $\chi^2$ fit to the masses 
and radii of about 100 spherical nuclei from $^{38}$Ca to $^{220}$Th with 
the result $a^{\rm s}_+ =0.600$, $h^{\rm s}_\nabla = 0.440$ and 
$h_{\rm Coul} = 0.941$, the rms deviations being 1.2 MeV and 0.01 fm for 
masses and radii, respectively (the zero-point energy correction 
$-\textstyle{3 \over 4}41A^{-1/3}$~MeV has been included). 
The LEDF in the suggested form with the above parameters has been 
abbreviated\cite{Fa98} as FaNDF$^0$. 

Typical results of the spherical LEDF calculations with FaNDF$^0$ 
are shown in Fig.~\ref{fig:pbsn} for even Pb and Sn isotopes, from the proton
to the neutron drip line, in comparison with experimental data and other 
model predictions. The ETFSI model is chosen as a reference. It is seen 
that the predictions obtained with the Gogny force just outside the 
measured regions are in strong disagreement with other models; the RMF 
calculations\cite{LRR99} yield larger oscillations in the masses around 
experimental values and seem to predict different behavior in the 
neutron-rich domain compared to others (Ref.\cite{LRR99} provides no results 
for the near-drip-line isotopes above $A=252$ for lead and above $A=156$ for 
tin). Approaching the neutron drip line, the FaNDF$^0$ masses fall in between 
the MM and ETFSI predictions.

The role of the term $\propto\!h_{\rm Coul}$ entering Eq.~(\ref{eCoul}) can be 
illustrated by the following example. The mass differences for the mirror 
pairs $^{17}$F--$^{17}$O and $^{41}$Sc--$^{41}$Ca calculated with FaNDF$^0$ are 
3.546 MeV and 7.174 MeV, respectively, whereas the corresponding experimental 
values are 3.543 MeV and 7.278 MeV. If one sets $h_{\rm Coul}=0$, the 
calculated mass differences for these mirror pairs would be respectively 
3.300 MeV and 6.872 MeV leading to a 6-7\% discrepancy known as the 
Nolen-Schiffer anomaly. It follows that Coulomb-nuclear correlations play 
an important role in finite nuclei. Incorporating the corresponding term 
in the LEDF improves the description of nuclear ground state properties and 
greatly reduces the severity of the Nolen-Schiffer anomaly.

\section{Variational principle and coordinate-space technique}
The LEDF calculations discussed above were performed by using the diagonal 
approximation in the pairing channel, with the particle continuum
approximated by the quasistationary states, with each wave function being 
normalized in a finite volume which includes only the decaying tail
beyond the nuclear surface. Such a HF+BCS technique eliminates to a great 
extent the presence of the ``particle gas'' and allows faster large-scale 
calculations. In certain situations, however, a more accurate 
treatment of the coupling with the particle continuum is needed (strongly 
surface-peaked pairing, nuclei near the neutron drip line, fine 
peculiarities in the behavior of differential observables, e.g.\ the 
staggering in charge radii, etc.). An appropriate tool is then the 
coordinate-space Gor'kov (or HFB) equations which naturally yield the 
localized ground state wave functions for finite systems, with correct 
asymptotics for the normal and anomalous 
densities\cite{Bul80,DFT84,BSTF87}. A more rigorous formulation
of the LEDF approach based on the general variational principle and the
coordinate-space technique is also of relevance\cite{FTTZ98}. In brief, 
we proceed as follows.

Generally, the energy $E$ of a nucleus with pairing is a functional of
the generalized density matrix $\widehat R$ which contains both a normal
component $\hat \rho$ and anomalous component $\hat \nu$:
\begin{equation}
E[\widehat R] = E_{\rm kin}[{\hat\rho}] +
E_{\rm int}[{\hat\rho},{\hat\nu}]\,,
\label{EHFB}
\end{equation}
where
$ E_{\rm kin}[{\hat\rho}] = {\rm Tr}(t\hat \rho)\,, $ and
$ E_{\rm int}[{\hat\rho},{\hat\nu}]
= E_{\rm int(normal)}[{\hat\rho}] +
E_{\rm anomal}[{\hat\rho},{\hat\nu}]$. The anomalous energy $E_{\rm
anomal}$ is chosen such that it vanishes in the limit $\nu \to 0$. In
the weak pairing approximation 
$\vert\Delta_{\rm F}/\epsilon_{\rm F}\vert\ll1$, 
which is usually the case for nuclear systems, 
one needs to retain only the first-order term $\sim\nu^2$: 
\begin{equation}
E_{\rm anomal}[{\hat\rho},{\hat\nu}]
= \frac 1 4 \left({\hat \nu}^\dagger{\hat{\cal F}}^{pp}_{\rm a}
[{\hat\rho}]{\hat\nu}\right)=\frac 1 2 \left({\hat\nu}^\dagger
\hat\Delta\right)\,, \label{Eanomal}
\end{equation}
where ${\hat{\cal F}}^{pp}_{\rm a}$ is the antisymmetrized
effective interaction in the pp channel and the 
parentheses imply integration over all variables. To calculate the 
ground state properties, one can now use the general variational 
principle with two constraints, 
$\langle{\rm HFB}\vert{\hat N(\mu)}\vert{\rm HFB}\rangle\equiv N(\mu)= 
N$ and ${\widehat R}^2 = \widehat R$,
leading to the variational functional of the form 
\begin{equation}
I[\widehat R] = E[\widehat R] - \mu N(\mu) - {\rm Tr}{\hat 
\Lambda}(\widehat R - {\widehat R}^2)\,, \label{VFUNC}
\end{equation} 
where $N$ is the particle number, $\mu$ the chemical potential, and
${\hat\Lambda}$ the matrix of Lagrange parameters 
(see e.g.\ Ref.\cite{RS80}). In general, 
the gap equation is nonlocal and its in-medium solution poses serious 
problems. We use the renormalization procedure\cite{FKTTZ} by 
introducing an arbitrary cutoff $\epsilon_{\rm c}$ in the energy space, but 
such that $\epsilon_{\rm c} > \epsilon_{\rm F}$, and by splitting the 
generalized density matrix into two parts, 
$\widehat R={\widehat R}_{\rm c}+\delta_{\rm c}{\widehat R}$, where 
$\delta_{\rm c}{\widehat R}$ is related to the integration over energies 
$\vert\epsilon\vert > \epsilon_{\rm c}$. The gap equation is renormalized 
to yield
\begin{equation}
\hat\Delta = \frac 1 2 \hat{\cal F}^\xi_{\rm a}{\hat\nu}_{\rm c}\,,
\label{rengap}
\end{equation}
where ${\hat\nu}_{\rm c}$ is the cutoff anomalous density matrix and
$\hat{\cal F}^\xi_{\rm a}$ is the effective antisymmetrized pairing 
interaction in which the contribution coming from the energy region 
$\vert\epsilon\vert > \epsilon_{\rm c}$ is included by renormalization.
With the Green's function formalism, for homogeneous infinite matter, 
it is shown that the variational function $E-\mu N$ does not change 
in first order in ${\hat\Delta}^2$ upon variation with respect to 
$\delta_{\rm c}{\widehat R}$. The total energy of the system and the chemical 
potential also remain the same if one imposes the above constraint 
$N(\mu)=N$ for the cutoff functional. To a good approximation, as 
discussed in Ref.\cite{FKTTZ}, this should be also valid for finite 
(heavy) nuclei. This result is expected since the major pairing
effects are developed near the Fermi surface and the pairing energy is
defined by a sum concentrated near the Fermi surface.
In infinite matter the pairing energy per particle is $E_{\rm pair}/N=
- 3\Delta^2_{\rm F}/8\epsilon_{\rm F}$.  It follows that, with the
cutoff functional, provided $\epsilon_{\rm c} > \epsilon_{\rm F}$, this 
leading pairing contribution is exactly accounted for. Thus the nuclear 
ground state properties can be described by applying the general 
variational principle to minimize the cutoff functional which has 
exactly the same form as~(\ref{VFUNC}) but with $\widehat R$ replaced by 
$\widehat R_{\rm c}$. Recalling now the Hohenberg-Kohn existence theorem, 
one can choose the energy functional to be of a local form, the LEDF, 
dependent on the normal and anomalous local densities $\rho(\vec r)$ and 
$\nu(\vec r)$. Then, after separating the spin variables, the anomalous 
energy density acquires the form of Eq.~(\ref{eanomal}), and~(\ref{rengap})
becomes simply the multiplicative gap equation
\begin{equation}
\Delta(\vec r) = {\cal F}^\xi(\vec r; [\rho_{\rm c}]) \nu_{\rm c}(\vec r)\,.
\label{gapeq}
\end{equation}
Having found the pairing and the HF potentials, the Gor'kov equations 
may be solved exactly with the coordinate-space 
technique\cite{BSTF87,FTTZ98}. The Green's functions obtained this way
are integrated in the complex energy plane to yield both the normal and
anomalous densities which are used then to compute the energy of the 
system. It is worthwhile to notice that our approach does not imply a 
cutoff of the basis since the general variational principle is 
formulated with a ``cutoff`` LEDF from which the ground state 
characteristics of a superfluid system may be calculated by using the 
generalized Green's function expressed through the solutions of the 
Bogolyubov equations at the stationary point. To construct the 
densities that appear in this local functional, only those Bogolyubov 
solutions from the whole set are needed which correspond to the 
eigenenergies $E_\alpha$ of the HFB Hamiltonian (which is a matrix of 
the first variational derivatives of the LEDF) up to the cutoff 
$\epsilon_{\rm c} > \epsilon_{\rm F}$. 

From the Gor'kov equations, after separating angular variables, for 
the generalized radial Green's function ${\hat g}_{jl}$ one gets the
equation
\begin{equation} \left( \begin{array}{cc}
 \epsilon-h_{jl}+\mu & -\Delta \\ -\Delta &  \epsilon+h_{jl}-\mu
\end{array} \right) \hat{g}_{jl}(r_1,r_2;\epsilon)= \left(
		   \begin{array}{cc} \delta(r_1-r_2) & 0 \\
 0 & \delta(r_1-r_2) \end{array} \right) \,,\label{RGF} 
\end{equation}
where $h_{jl}$ is the single-quasiparticle HF Hamiltonian in the $jl$
channel. The solution of this matrix equation can be constructed by
using the set of four linearly independent solutions which satisfy
the homogeneous (Bogolyubov) system of equations obtained 
from~(\ref{RGF}) by setting the right hand side to zero and which obey 
the physical boundary conditions both for the bound and scattering 
states\cite{BSTF87,FTTZ98}.

\section{Nuclear isotope shifts}
The formulated approach has been applied to the combined analysis 
of the differential observables such as neutron separation energies 
and odd-even effects in charge radii along isotopic chains. Reproducing
the observed changes of geometrical characteristics of nuclei, first 
of all the staggering and kinks in charge radii, could shed light on 
the density dependence of the effective pairing 
interaction\cite{ZRS87}. This point may be illustrated by considering 
how the density changes when the pairing gap appears in nuclear matter. 
To leading order, one finds the following expression for the energy per 
nucleon near the saturation point (henceforth, $x\equiv x_+$):
\begin{equation}
\frac E A = \frac{E_0}{A}+\frac{K_0}{18}\frac{(x-x_0)^2}{x_0^2}
+\beta(x)I^2-\frac 3 8 \frac{\Delta^2(x)}{\epsilon_{\rm F}(x)}\,
\label{Eaexp}
\end{equation}
where $K_0$ is the compression modulus and $\beta(x)I^2$ is the asymmetry 
energy with $I=(\rho_{\rm n}-\rho_{\rm p})/2\rho_0x \equiv (N-Z)/A$.
Due to the pairing term $\propto\!\Delta^2$, the position of the 
equilibrium point may be shifted to lower or higher densities 
depending on the behavior of $\Delta(x)$ near $x=1$. If $\Delta$
does not depend on $x$ in the vicinity of $x=1$ then the equilibrium 
density decreases due to presence of 
$\epsilon_{\rm F}(x)\!\propto\!x^{2/3}$ 
in the denominator of the pairing term (the system 
gains more binding energy). This means an expansion of the system. 
The effect is enhanced if $\Delta$ becomes larger during such an 
expansion, i.e. when the derivative $d\Delta(x)/dx$ at $x=1$ is 
negative. At equilibrium the pressure 
$P=\frac{\partial}{\partial x}(E/A) = 0$. For saturated symmetric 
nuclear matter without pairing the dimensionless density is, by 
definition, $x_0 = 1$. If $I \ne 0$ and $\Delta \ne 0$, the new 
equilibrium density $\rho=2\rho_0x$ can be found from the equation
\begin{equation}
x-x_0 =
\frac {9x_0^2}{K_0}\left[\frac {\Delta(x)} {4\epsilon_{\rm F}(x)}
\left(3\frac {d\Delta(x)}{dx}-\frac {\Delta(x)}{x} \right) - \frac
{d\beta(x)}{dx} I^2 \right]\,. \label{infrho}
\end{equation}
One can see that the density should be sensitive 
indeed to the derivatives of $\Delta$, and a negative slope in 
$\Delta$ should cause a decrease of the density. The obtained 
relations can be used to estimate the influence of pairing on the 
charge radii for heavy nuclei\cite{FTTZ94}. It was shown that pairing 
interaction with strong $\rho$-dependence at $x\approx 1$ might 
significantly change the equilibrium density. The size of this effect 
is controlled by the parameter 
$h^\xi$. For reasonable values of $h^\xi$, as given by Eq.~(\ref{pppar}) 
below, the shift of the saturation point is relatively small: 
$\vert\delta\rho\vert/2\rho_0 \leq 0.8$~\%. In finite nuclei, 
the surface term $\propto\!f^\xi_\nabla$ is equally important to 
produce a kink at magic neutron numbers and, especially, to explain 
the observed odd-even staggering in 
$\langle r^2 \rangle_{\rm ch}$ along isotopic chains\cite{FZ96}. 
The density dependence of the pairing force leads to the direct 
coupling between the neutron anomalous density $\nu_{\rm n}$ and the 
proton mean field. The suppression of $\nu_{\rm n}$ in 
the odd neutron subsystem because of the blocking effect influences 
the proton potential through the volume, $\propto\!h^\xi$, 
and surface, $\propto\!f^\xi_\nabla$, couplings, and this moves the 
behavior of the proton radii towards the desired regime\cite{FZ96}.
\begin{figure}[t]
\begin{center}
\epsfxsize=20pc 
\epsfbox{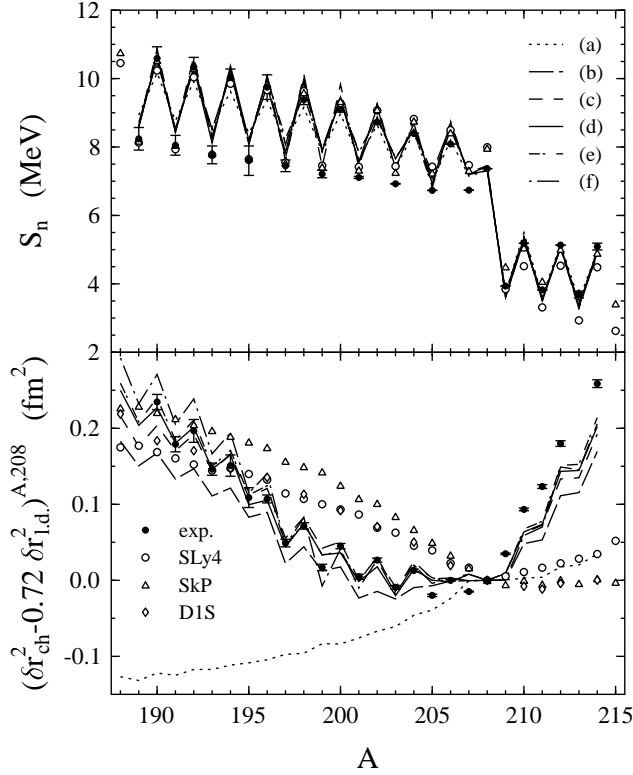}
\end{center}
\caption{Upper panel: neutron separation energies $S_{\rm n}$ for lead isotopes. 
Lower panel: differences of mean squared charge radii 
$\delta\langle r^2\rangle_{\rm ch}$ with respect to $^{208}$Pb as reference 
nucleus; 72\% of the corresponding liquid drop values (using $r_0=1.1$~fm) 
are subtracted to enhance the visibility of the small differences. 
Curves (a)--(f) correspond to the parameter sets (a)--(f) of the pairing 
force~(\ref{pppar}), respectively. Experimental data for $S_{\rm n}$
are from Ref.$^{16}$,
for $\delta\langle r^2\rangle_{\rm ch}$ from Refs.$^{26-28}$.
For comparison, the HFB calculations with Skyrme forces SkP
and SLy4, and with the Gogny D1S force but for even isotopes 
and $^{211}$Pb only, are also shown. \label{fig:pbsndr}}
\end{figure}
The results for the lead chain obtained with the elaborated 
coordinate-space technique and with the LEDF parametrization 
DF3 from Ref.\cite{BFKZ96} are shown in Fig.~\ref{fig:pbsndr}. 
The following 
parameter sets of the pairing force~(\ref{fxi}) are deduced 
(for the energy cutoff $\epsilon_{\rm c}=40$~MeV and the exponent $q=2/3$):
\begin{equation}
\begin{array}{lllllllr}
  f^\xi_{\rm ex}=-0.56,&h^\xi=0,   &f^\xi_\nabla=0  &\; (a); &
\;f^\xi_{\rm ex}=-1.20,&h^\xi=0.56,&f^\xi_\nabla=2.4&\; (b);  \\
  f^\xi_{\rm ex}=-1.60,&h^\xi=1.10,&f^\xi_\nabla=2.0&\; (c); &
\;f^\xi_{\rm ex}=-1.79,&h^\xi=1.36,&f^\xi_\nabla=2.0&\; (d);  \\
  f^\xi_{\rm ex}=-2.00,&h^\xi=1.62,&f^\xi_\nabla=2.0&\; (e); &
\;f^\xi_{\rm ex}=-2.40,&h^\xi=2.16,&f^\xi_\nabla=2.0&\; (f).  \\
\end{array}\,
\label{pppar}
\end{equation}
As seen in the upper panel of Fig.~\ref{fig:pbsndr}, the neutron 
separation energies $S_{\rm n}$ are reproduced equally well for all
parameter sets (a)--(f) of Eq.~(\ref{pppar}). The HFB calculations with 
two state-of-the-art Skyrme functionals SkP (Ref.\cite{DFT84}) and 
SLy4 (Ref.\cite{SLy4}) describe the $S_{\rm n}$ values with more or 
less the same 
quality as our curves (a)--(f), some deviations are observed above 
$^{207}$Pb. In the lower panel the calculated isotope shifts of mean 
squared charge radii, $\delta\langle r^2\rangle_{\rm ch}$,  
with respect to $^{208}$Pb as a reference nucleus are presented. Our 
curves except (a) do not differ from each other significantly and all
of them reproduce qualitatively the kink and the size of 
odd-even staggering. The curve (a) corresponds to a simple contact 
pairing force which does not depend on density; in this case the 
behavior of $\delta\langle r^2\rangle_{\rm ch}$ is rather smooth and 
neither kink nor staggering are reproduced. One can see also that 
the Skyrme functionals SkP and SLy4 give too small kink and too small
staggering. The Gogny D1S force could not reproduce the kink 
either. This fact clearly demonstrates the importance of including
odd-mass nuclei in the fitting procedure of the force constants.

\section{Ground-state properties of nuclear matter}
Now, the empirical information gained from finite laboratory nuclei 
can be used to study the ground state properties and behavior of 
$\Delta$ as a function of density $\rho=2\rho_0x$ (or the Fermi 
momentum $k_{\rm F}=(3\pi^2\rho/2)^{1/3}=k_{0\rm F}x^{1/3}$) in
infinite nuclear matter. Since the density-gradient term 
$\propto\!f^\xi_\nabla$ vanishes in this case, only two parameters 
are relevant: $f^\xi_{\rm ex}$ and $h^\xi$. The gap 
equation~(\ref{gapeq}) reduces to
\begin{equation}
\Delta(x)=-
\int\nolimits_{k\le k_{\rm c}}\,\frac{d\vec k}{(2\pi)^3}{\cal F}^{\xi}(x)
\frac{\Delta(x)}{2\sqrt{(\epsilon_k-\epsilon_{\rm F}(x))^2
+\Delta^2(x)}}\,, \label{gapinf}
\end{equation}
where $k_{\rm c} = \sqrt{2m(\epsilon_{\rm F}+\epsilon_{\rm c})}/\hbar$ and
$\epsilon_k=\hbar^2k^2/2m$. Its solution in the weak pairing regime
can be written as\cite{Fa99}:  
\begin{equation} 
\Delta(k_{\rm F})=c\epsilon_{\rm F}\exp 
\left[-\frac{\pi}{2}\cot\delta(k_{\rm F})\right]\,, \label{delcot} 
\end{equation} 
where $c=8{\rm e}^{-2}\approx 1.083$ and where
we have introduced the Fermi level phase shift 
$\delta(k_{\rm F})$ defined by 
\begin{equation} 
k_{\rm F}\cot\delta(k_{\rm 
F}) = -\frac{4k_{0\rm F}}{\pi}\left(\frac{1}{f^\xi(k_{\rm F})}+ 
\frac{k_{\rm c}(k_{\rm F})}{2k_{0{\rm F}}}\right)
-\frac{k_{\rm F}}{\pi}\ln\left(\frac{k_{\rm c}(k_{\rm F})-k_{\rm F}}
{k_{\rm c}(k_{\rm F})+k_{\rm F}}\right)\,, \label{kcot}
\end{equation} 
with $k_{\rm c}(k_{\rm F})=\sqrt{k_{0c}^2+k_{\rm F}^2}$; 
$k_{0c}=\sqrt{2m\epsilon_{\rm c}}/\hbar$. Eq.~(\ref{kcot}) corresponds to 
an exact solution of the nn scattering problem at the relative 
momentum $k=k_{\rm F}$ with the states truncated by a momentum cutoff 
$k_{\rm c}=k_{\rm c}(k_{\rm F})$ for contact potential 
$C_0f^\xi(k_{\rm F})\delta({\vec r})$ (see, e.g., Ref.\cite{EBH97}). 

\begin{figure}[t]
\begin{center}
\epsfxsize=20pc
\epsfbox{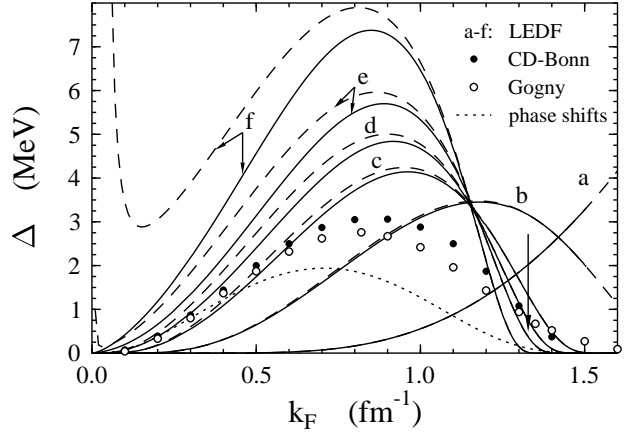}
\end{center}
\caption{Pairing gap in nuclear matter as a function of the Fermi momentum. 
Full (dashed) curves (a)--(f) are obtained by solving Eqs.~(\ref{gapinf}) and 
(\ref{partnum}) (by using the weak pairing approximation of 
Eqs.~(\ref{delcot}) and~(\ref{kcot})) 
with contact pairing force~(\ref{fxi}) and energy cutoff 
$\epsilon_{\rm c}=40$~MeV, and correspond, respectively, to the parameter sets 
(a)--(f) of Eq.~(\ref{pppar}). The vertical arrow marks 
the Fermi momentum at saturation point for the functional DF3 without 
pairing, $k_{0\rm F}=1.328$~fm$^{-1}$. The solid (open) circles are the 
solutions of the nonlocal gap equation with the CD-Bonn 
potential$^{32}$                                         
(with the finite-range Gogny D1 force$^{33}$).           
The dotted line is obtained from 
Eq.~(\ref{delcot}) with the free NN scattering phase shifts 
(see text). \label{fig:dnm}}
\end{figure}

Shown in Fig.~\ref{fig:dnm} are the results for $\Delta$ in nuclear 
matter with parameter sets of Eq.~(\ref{pppar}). Interestingly, for 
the sets (b)--(e), which reproduce satisfactorily both the $S_{\rm n}$ 
and $\delta\langle r^2 \rangle_{\rm ch}$ values, there
exists a ``pivoting'' point at $k_{\rm F}\approx 1.15$~fm$^{-1}$
(at $\approx$0.65 of the equilibrium density) with the same value of
$\Delta_{\rm piv}\approx 3.3$~MeV. The approximation~(\ref{delcot}) 
(dashed curves in Fig.~\ref{fig:dnm}) works well in the entire range of 
$k_{\rm F}$ for the set (a), but for the other sets this is true only 
at $k_{\rm F}$ greater than $\approx$1.2~fm$^{-1}$ and also, for the 
sets (b), (c) and (d), at $k_{\rm F}$ less than 0.42, 0.14 and 
0.042~fm$^{-1}$, respectively (in these regions, 
$\Delta/\epsilon_{\rm F}\leq 0.1$). It should be stressed that 
$\epsilon_{\rm F}$ entering the integrand of Eq.~(\ref{gapinf}) can be 
expressed directly through density $\rho$ by 
$\epsilon_{\rm F}=\hbar^2k_{\rm F}^2(\rho)/2m$ with 
$k_{\rm F}(\rho)=(3\pi^2\rho/2)^{1/3}$ only if the pairing is weak and 
the dependence of the Fermi energy $\epsilon_{\rm F}$ (and the chemical 
potential $\mu$) on $\Delta$ can be disregarded. Otherwise one should 
introduce the particle number condition
\begin{equation}
x=\frac{2}{\rho_0}\int\nolimits_{k\le k_{\rm c}}\,\frac{d\vec
k}{(2\pi)^3}n_k(x),\quad n_k(x)= \frac 1 2
\left(1-\frac{\epsilon_k-\epsilon_{\rm F}(x)}
{\sqrt{(\epsilon_k-\epsilon_{\rm F}(x))^2+\Delta^2(x)}}\right)\,,
\label{partnum}
\end{equation}
and to solve the system of the two equations~(\ref{gapinf}) 
and~(\ref{partnum}) with respect to $\Delta$ and $\epsilon_{\rm F}$. 
The results shown in Fig.~\ref{fig:dnm} by the solid curves correspond 
to such a solution.
\begin{figure}[t]
\begin{center}
\epsfxsize=18pc
\epsfbox{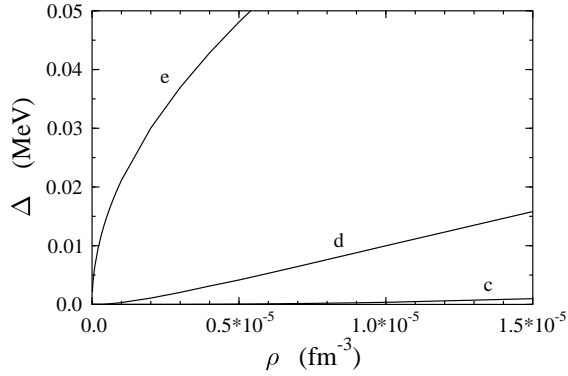}
\end{center}
\caption{Pairing gap $\Delta$ as a function of $\rho$ at very low densities. 
Curves (c)--(e) correspond to the parameter sets (c)--(e) of Eq.~(\ref{pppar}),
respectively. \label{fig:dellow}}
\end{figure}
All parameter sets~(\ref{pppar}) except (a) reproduce the neutron separation
energies and the isotope shifts of charge radii $\langle r^2 \rangle_{\rm ch}$ 
of lead isotopes fairly well (see Fig.~\ref{fig:pbsndr}). Shown also in 
Fig.~\ref{fig:dnm} are the values of the $^1S_0$ pairing gap in nuclear 
matter obtained for the CD-Bonn potential\cite{EHJ98} and for the Gogny D1 
force\cite{Kuch89}. The agreement between the two latter calculations is 
relatively good, while both deviate noticeably from our predictions. The 
curve for density-independent force, set (a), stands by itself with a 
positive derivative $d\Delta(x)/dx$ everywhere; no acceptable description of 
$\langle r^2 \rangle_{\rm ch}$ could be obtained in this case\cite{FTTZ98}.  

At very low densities (\ref{kcot}) reduces to
\begin{equation}
k_{\rm F}\cot\delta(k_{\rm F}) \approx -\frac{1}{a_{\rm nn}} + 
\frac{1}{2}r_{\rm nn}k_{\rm F}^2 - \frac{2k_{\rm F}}{\pi}
\left[\frac{k_{\rm F}}{2k_{0c}}-\frac{2h^\xi}{(f^\xi_{\rm ex})^2}
\left(\frac{k_{\rm F}}{k_{0\rm F}}\right)^{3q-1}\right]\,,
\label{kcot0}
\end{equation}
where $a_{\rm nn}$ is the singlet scattering length, 
\begin{equation}
a_{\rm nn}= \frac{\pi}{2k_{0\rm F}}
\left(\frac{\sqrt{2m\epsilon_{\rm c}}}{\hbar k_{0\rm F}}+
\frac{2}{f^\xi_{\rm ex}}\right)^{-1} \equiv
\frac{\pi}{4k_{0\rm F}}\left(\frac{1}{f^\xi_{\rm ex}}-
\frac{1}{f^\xi_{\rm cr}}\right)^{-1}\,,
\label{scl}
\end{equation}
and $r_{\rm nn}$ is the effective range, $r_{\rm nn}=4/\pi k_{0c}$. Here we have 
introduced the critical constant $f^\xi_{\rm cr}=-2k_{0\rm F}/k_{0c}$,  
the vacuum strength $f^\xi_{\rm ex}$ at which the two-nucleon 
problem has a bound state solution at zero energy (in our case, 
$f^\xi_{\rm cr}=-1.912$).  

The first two terms in~(\ref{kcot0}) would describe the $s$-wave 
phase shift at low energies through an expansion of $k\cot\delta$ in 
powers of the relative momentum $k=k_{\rm F}$ if the interaction were 
density-independent -- i.e., if the coupling strength and momentum 
cutoff were fixed by $f^\xi=f^\xi_{\rm ex}$ and $k_{\rm c}=k_{0c}$, respectively. 
It follows that, with a density-dependent effective force, such an expansion 
contains additional terms which are, for the parametrization used here, 
of the same order as the effective range term. This simply 
demonstrates that, for reproducing the pairing gap, the effective interaction 
even at very low densities need not necessarily coincide with the bare NN 
interaction as was discussed by Migdal many years ago\cite{Migdal}.

At very low densities, at $k_{\rm F}\rightarrow 0$, to leading order from 
(\ref{kcot}) we obtain
\begin{equation}
\Delta=
c\epsilon_{\rm F}\exp{\left(\frac{\pi}{2k_{\rm F}a_{\rm nn}}\right)}\,,
\quad a_{\rm nn}<0\,.
\label{gapweak}
\end{equation}
This expression agrees with the results of Ref.\cite{KKC96} based on a 
general analysis of the gap equation at low densities\footnote{Our effective 
pairing interaction with the choice $q=1/3$ would lead in the dilute limit to 
the expression for $\Delta$ of the form of Eq.~(\ref{gapweak}) but with a 
different prefactor $c$ depending on the value of $h^\xi$. If, furthermore, 
we define it by $h^\xi=(1+2\ln 2)(f^\xi_{\rm ex})^2/6$ we get in the leading
order $\Delta(k_{\rm F})=(2/e)^{7/3}\epsilon_{\rm F}\exp(\pi/2k_{\rm F}a)$,
i.e.\ the result obtained in Ref.\cite{GMB61} for a non-ideal Fermi gas 
with taking into account the terms up to the second order in $k_{\rm F}a$.}
when $k_{\rm F}\vert a_{\rm nn} \vert \ll 1$. 

But we should stress that~(\ref{gapweak}) is valid only 
in the weak-coupling regime corresponding to negative $a_{\rm nn}$. In the 
opposite case, the gap in the dilute limit has to be found in a different 
way. At $f^\xi_{\rm ex} > f^\xi_{\rm cr}$, from (\ref{gapweak}) and~(\ref{delcot}) 
it follows that at low densities the pairing gap is exponentially small and 
eventually $\Delta(k_{\rm F}\rightarrow 0)=0$. Such a weak pairing regime 
with Cooper pairs forming in a spin singlet $l=0$ state exists up to the 
critical point at which the attraction becomes strong enough to change the 
sign of the scattering length. Then the strong pairing regime sets in, and 
$\Delta$ should be determined directly from the combined solution of 
Eqs.~(\ref{gapinf}) and~(\ref{partnum}). In the dilute systems, 
$\epsilon_{\rm F}$ plays the role of the chemical potential $\mu$. 
At the critical point, $\mu$ becomes negative, and a bound 
state of a single pair of nucleons with the binding energy $\epsilon_b=2\mu$
($=-\hbar^2/ma^2_{\rm nn}$) becomes possible\cite{NSR85}. In this regime,
$\Delta$ can be found from (\ref{partnum}). In the leading order we get
\begin{equation}
\Delta=\frac{\hbar^2}{m}\left(\frac{2\pi\rho}{a_{\rm nn}}\right)^{1/2}\,,
\quad a_{\rm nn}>0\,.
\label{gapstrong}
\end{equation}
It follows that, in the dilute case, the energy needed to break a condensed 
pair goes smoothly from $2\Delta$ to $2\mu=\epsilon_b$ as a function of the 
coupling strength as the regime changes from weak to strong pairing. But as 
seen from~(\ref{gapweak}) and~(\ref{gapstrong}), the behavior of $\Delta$  
is such that the derivative $d\Delta/d\rho$ at $\rho\rightarrow 0$ as a 
function of $f^\xi_{\rm ex}$ exhibits a discontinuity from 0 to $\infty$.  
This is illustrated in Fig.~\ref{fig:dellow} where we have plotted 
$\Delta(\rho)$ for the sets (c)--(e) of Eq.~(\ref{pppar}), which embrace 
both regimes. Notice also that both analytical expressions~(\ref{gapweak}) 
and~(\ref{gapstrong}) give a pure imaginary gap at the critical point,
where the scattering length changes sign.
When the Fermi momentum $k_{\rm F}$ approaches from below the upper 
critical point, where the pairing gap closes, $\Delta$ becomes exponentially 
small\cite{KKC96}. In weak coupling, $\Delta$ is also exponentially small 
at low densities. It is noteworthy that, as follows from~(\ref{delcot}), in 
both these cases the gap closes exactly at the points where the phase shift 
passes zero. As an illustration, we show in Fig.~\ref{fig:dnm} by the 
dotted line the values of $\Delta(k_{\rm F})$ obtained from~(\ref{delcot}) 
using the ``experimental'' nn phase shifts\footnote{We thank Rupert 
Machleidt for providing us with these nn phase shifts.}. It is 
seen that $\Delta$ obtained this way closely follows the solution of the gap 
equation with the CD-Bonn potential\cite{EHJ98} at low densities. The nn 
phase shift passes zero at the relative momentum $k=1.71$~fm$^{-1}$, and the 
gap should vanish at the corresponding Fermi momentum. Unfortunately, the 
solutions for $\Delta$ are given in Ref.\cite{EHJ98} only in the region up 
to $k_{\rm F}=1.4$~fm$^{-1}$.
\begin{figure}[b]
\begin{center}
\epsfxsize=20pc 
\epsfbox{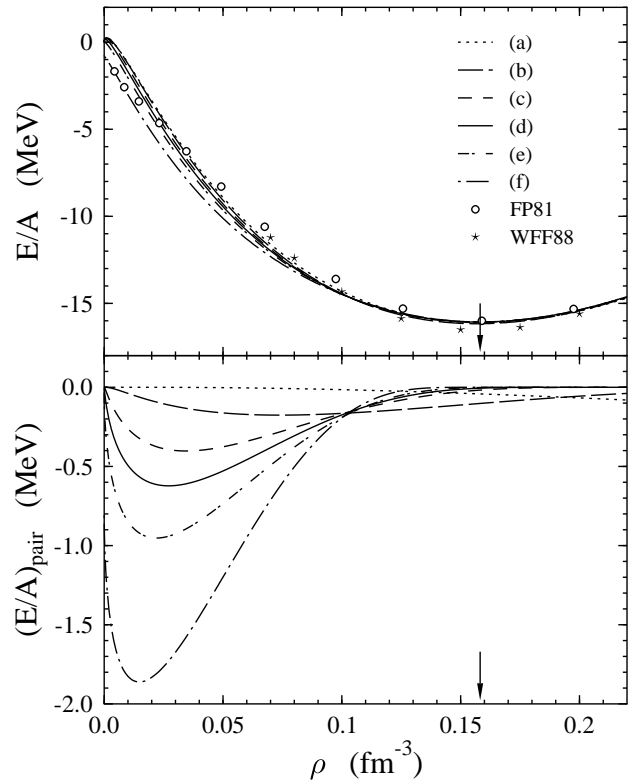}
\end{center}
\caption{Energy per nucleon $E/A$ (top) and pairing contribution to $E/A$
(bottom) in nuclear matter. Curves (a)--(e) are calculated using 
Eq.~(\ref{EOS}) and correspond to the strength parameters (a)--(e) of 
Eq.~(\ref{pppar}), respectively. Open circles and stars are the 
calculations of Ref.$^{10}$                            
and Ref.$^{11}$,                                       
respectively, for the UV14 plus TNI model. The vertical arrows mark the 
saturation density for the functional DF3 without pairing, 
$2\rho_0=0.1582$~fm$^{-3}$. \label{fig:eanm}}
\end{figure}
\begin{figure}[t]
\begin{center}
\epsfxsize=18pc
\epsfbox{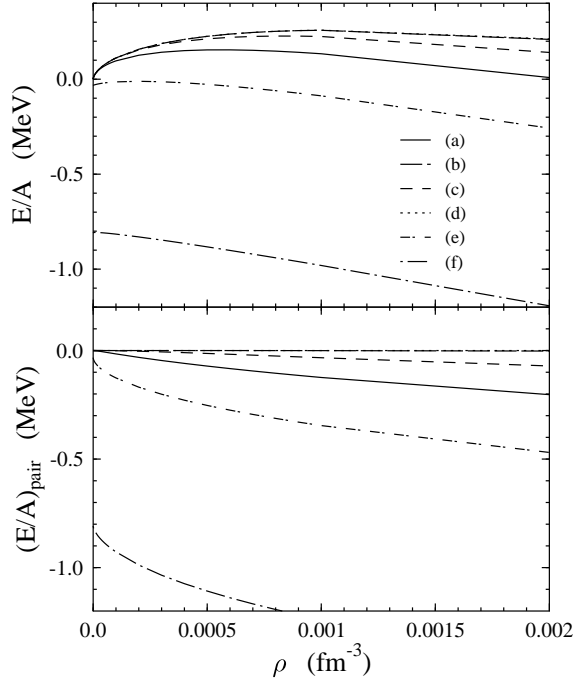}
\end{center}
\caption{Energy per nucleon $E/A$ (top) and pairing contribution to
$E/A$ (bottom) for nuclear matter at low densities. The notations 
are the same as in Fig.~\ref{fig:eanm}. \label{fig:ealow}}
\end{figure}
For nuclear matter, with our LEDF, the energy per particle is 
\begin{equation} 
\frac{E}{A}(x) = \frac{2}{\rho_0 x}\int\nolimits_{k\le k_{\rm c}} \frac{d\vec 
k}{(2\pi)^3}\frac{\hbar^2k^2}{2m}n_k(x)+ \frac{1}{3}\epsilon_{0\rm 
F}a_+^{\rm v}f_+^{\rm v}(x)x+ \frac{3\Delta^2(x)}{2f^\xi(x)x
\epsilon_{0\rm F}}\,, 
\label{EOS} 
\end{equation} 
where $f_+^{\rm v}(x)=(1-h^{\rm v}_{1+}x)/(1+h^{\rm v}_{2+}x)$. 
Numerically, $a_+^{\rm v}=-6.422$, 
$h^{\rm v}_{1+}=0.163$ and $h^{\rm v}_{2+}=0.724$ (Ref.\cite{BFKZ96}). 
In the dilute limit the ``particle-hole'' 
term $\propto\!f_+^{\rm v}$ vanishes  linearly in density. 
The chemical potential is 
\begin{equation} 
\mu(x)=\epsilon_{\rm 
F}(x)+\frac{1}{3}\epsilon_{0\rm F} a_+^{\rm v}[f^{{\rm v}\prime}_+(x)x^2+
2f^{\rm v}_+(x)x] 
+\frac{3f^{\xi\prime}(x)}{2f^{\xi 2}(x)}
\frac{\Delta^2(x)}{\epsilon_{0\rm F}}\,,
\label{chimpot}
\end{equation}
where the prime denotes a derivative with respect to $x$,  
$\epsilon_{\rm F}(x)$ and $\Delta(x)$ are determined from~(\ref{gapinf}) 
and~(\ref{partnum}). The two last terms in~(\ref{chimpot}), even in strong 
pairing regime, vanish in the dilute limit at least as $x^q$ if $0<q<1$ or 
linearly in $x$ if $q\geq 1$. We see again that, in strong coupling, 
in the leading order $\mu=\epsilon_{\rm F}=\epsilon_b/2 < 0$.  
The calculated energy per nucleon as a function of the isoscalar density
$\rho$ is shown in the upper panel in Fig.~\ref{fig:eanm} together with the 
results of the nuclear matter calculations\cite{FP81,WFF88}. It is seen 
that DF3 gives qualitatively reasonable description of the nuclear matter 
EOS and that pairing could contribute noticeably to the binding energy 
especially at lower densities. In the lower panel in Fig.~\ref{fig:eanm} 
we have plotted the pairing energy per nucleon, 
$(E/A)_{\rm pair}$, obtained by subtracting from (\ref{EOS}) the 
corresponding value of $E/A$ at $\Delta=0$. The pairing contribution 
increases, as expected, when $f^\xi_{\rm ex}$ becomes gradually more 
attractive, with a shift to lower densities. For the sets (e) and (f) the 
attraction is strong, $f^\xi_{\rm ex} < f^\xi_{\rm cr}$. In these cases a 
nonvanishing binding energy in the dilute limit is solely due to 
Bose-Einstein condensation of the bound pairs, the spin-zero bosons, when all 
the three quantities, $\mu$, $E/A$ and $(E/A)_{\rm pair}$, reach the same 
value $\epsilon_b/2=\hbar^2/2ma^2_{\rm nn}$ ($\epsilon_b = -0.0646$ and 
$-1.616$~MeV for the set (e) and (f), respectively). This is illustrated in 
Fig.~\ref{fig:ealow} where we have plotted $E/A$ and $(E/A)_{\rm pair}$ as 
functions of $\rho$ at very low densities.
 
We have considered the properties of nuclear matter with s-wave pairing 
within the LEDF framework, including extrapolation to 
the dilute limit, with a few possible parameter sets of the pairing force 
deduced from finite nuclei. At low densities, in the 
case of symmetric $N=Z$ matter, the $^3S_1-^3D_1$ pairing could be more 
important since the n--p force is more attractive than in the p--p or 
n--n pairing channels. Thus, our approach, with the $^1S_0$ pairing only, 
would be more appropriate for an asymmetric $N\neq Z$ case and for pure 
neutron systems. From this point of view the best choice for the LEDF 
calculations seems to be the parameter set (d) of Eq.~(\ref{pppar}) since 
it gives the singlet scattering length $a_{\rm nn}\approx -17.2$~fm which 
corresponds to a virtual state at $\approx 140$~keV known experimentally. 
As seen in Fig.~\ref{fig:dnm}, for this choice the behaviour of $\Delta$ 
at low densities agrees well with the calculations based on realistic NN 
forces. At higher densities, however, our predictions for $\Delta$ with the 
set (d) go much higher reaching a maximum of $\approx 4.84$~MeV at 
$k_{\rm F}\approx 0.92$~fm$^{-1}$, while the calculations of 
Ref.\cite{EHJ98} give a maximum of about 3~MeV at $k_{\rm 
F}\approx 0.82$~fm$^{-1}$. With a bare NN interaction, if one assumes 
charge independence and that $m_{\rm n}=m_{\rm p}$, the pairing 
gap would be, at a given $k_{\rm F}$, exactly the same both in symmetric 
nuclear matter and in neutron matter. As shown in Refs.\cite{WAP93,SCL96}, 
if one includes medium effects in the effective pairing interaction, $\Delta$ 
in neutron matter would be reduced substantially, to values of the order of 
1~MeV at the most. Whether such a mechanism works in the same direction for 
symmetric nuclear matter is still an open question. 
The force~(\ref{fxi}) was chosen to be dependent on 
the isoscalar density only, since we have analyzed the existing data 
on $S_{\rm n}$ and $\delta\langle r^2\rangle_{\rm ch}$ for finite nuclei with a 
relatively small asymmetry, $(N-Z)/A\leq 0.25$. An extrapolation to neutron 
matter with such a simple force would give a larger pairing gap than for 
nuclear matter. This suggests that some additional dependence on the 
isovector density $\rho_{\rm n}-\rho_{\rm p}$ might be present in the effective 
pairing interaction.

\section{Staggering and kinks in nuclear multipole excitations} 
The staggering phenomenon observed in the behavior of charge radii plotted 
as a function of neutron number is a prominent odd-even effect 
which has been systematically measured and widely discussed. Similar effects 
are expected to exist for other quantities such as neutron and matter radii, 
centroid energies of multipole excitations (position of the giant 
resonances), etc. 
\begin{figure}[b]
\begin{center}
\epsfxsize=18pc
\epsfbox{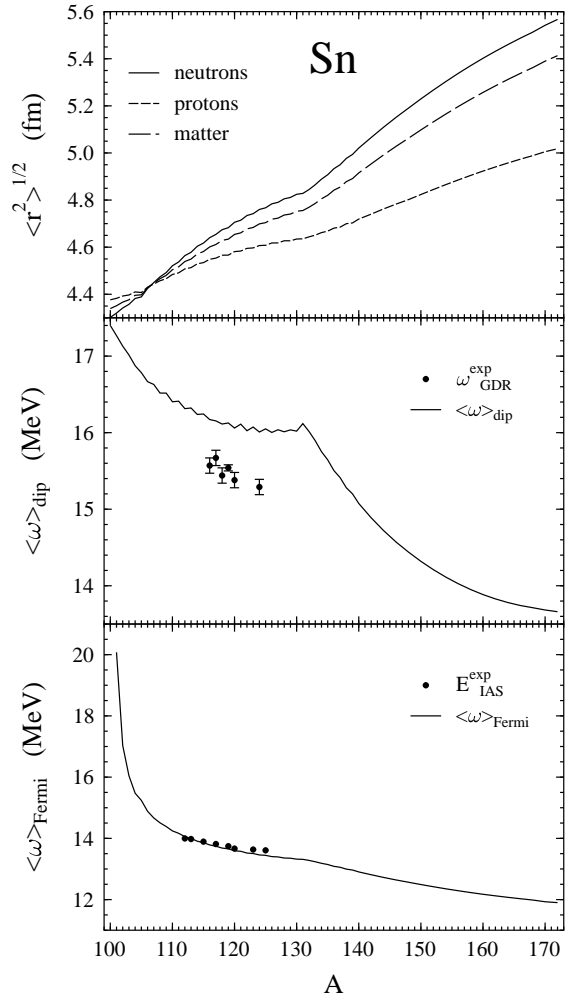}
\end{center}
\caption{Neutron, proton and matter radii (top), mean energy of isovector
dipole (middle) and Fermi charge-exchange (bottom) transitions for tin
isotopes. The solid dots for $\bar\omega_{\rm dip}$ correspond to the 
experimental data for the maximum of the dipole photoabsorption Lorentz 
curves$^{44}$,
while those for $\bar\omega_F$ to the experimental positions of the isobaric 
analog states with respect to the daughter nuclei$^{45}$.
\label{fig:gdrcul}}
\end{figure}
Shown in the upper panel of Fig.~\ref{fig:gdrcul} are the 
rms proton, neutron and matter radii for the tin isotopic chain calculated 
with the functional DF3 and with the set (d) of the pairing force. In the 
middle panel we show the mean energies $\bar\omega_{\rm dip}$
of dipole isovector excitations. These are calculated as the square root of 
the ratio $m_3/m_1$ with $m_1$ and $m_3$ the linear and cubic energy-weighted 
sum rule, i.e. the first and the third moment of the corresponding RPA 
strength distribution, respectively\cite{TF82}:
\begin{equation}
\bar\omega_{\rm dip}=\left[-{{\hbar^2}\over{3m}}
{A\over{NZ}} \int d\vec r d\vec r\,'\,
{{\partial\rho_{\rm n}}\over{\partial\vec r}}
{\cal F}^{\rm np}(\vec r,\vec r\,')
{{\partial\rho_{\rm p}}\over{\partial\vec r\,'}}
\right]^{1/2}\,.
\label{odip}
\end{equation}
Here  ${\cal F}^{\rm np}$ is the effective neutron--proton interaction 
obtained from the energy-density functional as the second variational 
derivative $\delta^2 E_{\rm int}/\delta\rho_{\rm n}\delta\rho_{\rm p}$. 
In the lower panel of Fig.~\ref{fig:gdrcul} we show the mean energies 
$\bar\omega_{\rm F}$ of the charge-exchange $0^+$ excitations, i.e. the 
Fermi transitions, in tin nuclei with $N>Z$. These energies are calculated 
within the sum rule approach as the ratio $(m_1^++m_1^-)/(m_0^+-m_0^-)$
with $m_1^+$ and $m_1^-$ the first moment of the strength 
distribution of the Fermi transitions in the $\beta^+$ and $\beta^-$ channel, 
respectively (energy-weighted sum rules) and with $m_0^+$ and $m_0^-$ 
the corresponding non-energy weighted sum rules. The expression 
for $\bar\omega_{\rm F}$ reads (see, e.g. Ref.\cite{PF83}): 
\begin{equation}
\bar\omega_{\rm F}={1\over{N-Z}}\int d\vec r\, U_{\rm Coul}(\vec r\,)
\left(\rho_{\rm n}(\vec r\,)-\rho_{\rm p}(\vec r\,)\right)\,,
\label{suru}
\end{equation}
where $U_{\rm Coul}$ is the Coulomb mean field potential.
It is seen in Fig.~\ref{fig:gdrcul} that the neutron, proton and matter radii 
as functions of the mass number $A$ reveal a kink at magic $^{132}$Sn, and at 
larger $A$ the difference between rms neutron and proton radii starts to 
increase more rapidly. The staggering in radii is observed mostly in the
region between the two magic nuclei, from $^{100}$Sn to $^{132}$Sn, and this
effect is practically washed out beyond $A=140$. The mean energy of dipole 
transitions occurs to be in anticorrelation with such a behavior in radii,
and this seem to be in qualitative agreement with experimental data (one
should mention that Eq.~(\ref{odip}) overestimates the position of the
giant dipole resonance by $\approx 0.5$~MeV because of the cubic 
energy-weighted sum rules used in the derivation of $\bar\omega_{\rm dip}$).
One also observes a distinct kink in the behaviour of $\bar\omega_{\rm dip}$ 
at $A=132$. Beyond this magic number the mean dipole energy decreases
rather fast and then nearly saturates when approaching $A=172$. This might 
be connected with an enhancement of the low-energy dipole 
transitions\cite{Ham96} and also with possible appearance of the so-called 
soft dipole mode\cite{Fay91} in nuclei near the neutron drip line. The 
anticorrelations in the staggering behavior of $\bar\omega_{\rm dip}$ and 
the rms radii $\langle r^2\rangle^{1/2}_{\rm n}$, 
$\langle r^2\rangle^{1/2}_{\rm p}$ can 
easily be understood by considering the influence of pairing on the gradients 
of neutron and proton densities entering Eq.~(\ref{odip}). The odd-even 
effect in $\bar\omega_{\rm dip}$ is constructive and more pronounced 
in the $A\leq 132$ region where both gradients strongly overlap. Beyond
$A = 132$ the larger differences between neutron and proton rms 
radii imply a lower overlap between neutron and proton density gradients 
at the nuclear surface, hence a smaller mean dipole energy. The situation
with mean energy of the Fermi charge-exchange transitions is different. 
From Eq.~(\ref{suru}) one expects that the correlation between $\bar\omega_F$
and the neutron and proton rms radii should be destructive. As
seen in the lower panel in Fig.~\ref{fig:gdrcul}, the staggering in the    
evolution of $\bar\omega_F$ with $A$ is very weak indeed and almost 
invisible. The kink at the magic mass number $A=132$ is also much less 
pronounced compared to the dipole case. Remarkable enough, the theoretical 
self-consistent sum-rule predictions for $\bar\omega_F$ in tin isotopes
are in excellent agreement with the available experimental data on the
position of the isobaric analog states.  

\section*{Acknowledgments}
It is a pleasure to thank our colleagues Sergei Tolokonnikov and Eugene
Trykov for their contribution to this work. Partial support by the Deutsche 
Forschungsgemeinschaft and by the Russian Foundation for Basic Research 
(project 98-02-16979) is gratefully acknowledged.

\end{document}